\documentclass{desyprocA4}
\renewcommand{\tt}[1]{\texttt{#1}}

\begin{document}
\title{WHIZARD @ LCFORUM 2012: A Status Report}

\author{{\slshape J\"urgen Reuter$^1$}\\[1ex]
$^1$DESY, Notketra{\ss}e 85, 22607 Hamburg, Germany}

\contribID{xy}

\confID{4980}  
\desyproc{DESY-PROC-2012-XX}
\acronym{LC2012} 
\doi  

\maketitle

\begin{abstract}
This is a status report of the WHIZARD Monte Carlo multi-purpose event
generator given at the LCFORUM 2012 at DESY. In case you use the
program, please do cite the official
reference(s),~\cite{Kilian:2007gr,Moretti:2001zz}. I review here the
development of the WHIZARD generator version 2 with a special emphasis
on linear collider physics.  
\end{abstract}

\section{Introduction}

The multi-purpose Monte Carlo event generator WHIZARD was developed as
a tool for linear collider physics during the late
1990s~\cite{Kilian:2001qz}. Some of the first studies with exclusive
four, six and eight fermion final states for linear collider physics
have been done with
WHIZARD~\cite{Kilian:2002cg,Kilian:2004sr,Boos:1997gw,Boos:1999kj}.
The first public version, 1.00, of WHIZARD has been released in December
2000. It was written in \texttt{Fortran90/95} and used from its
beginnings the VAMP package~\cite{Ohl:1998jn} for a multi-channel
adaptive Monte Carlo integration. The major improvement was an
algorithm to model the phase space channels for a process under
consideration and to provide the corresponding phase space mappings to
flatten out the divergencies of the integrand for an optimized
importance sampling. In the 1.xx (now called legacy) versions of
WHIZARD, matrix elements from early version of
MadGraph~\cite{Stelzer:1994ta} and CompHep~\cite{Boos:1994xb} as well
as from the at that time newly developed Optimized Matrix Element
Generator O'Mega~\cite{Moretti:2001zz} could be used. Parton shower
and hadronization could be simulated via an interface to
PYTHIA~\cite{Sjostrand:2001yu}. 

During the years 2001-2005/06, many technical and physics features
have been added on demand of either theoretical or experimental users
of the program or the authors itself. Support for several event file
formats have been added. For a realistic simulation of linear lepton
colliders, the ability to use structured beams have become crucial,
specifically for experimental feasibility studies and detector
development. Along these lines, initial state radiation (ISR)
following the approach of Ref.~\cite{Skrzypek:1990qs}, $k_T$
distribution of the radiating initial beams as well as explicit
photons from ISR in the final state events. Beamstrahlung, i.e. the
modifications of the beam spectra due to classical electromagnetic
interactions of the lepton beams, as well as photon beam options via
Compton backscattering off laser photons could be simulated by
attaching the CIRCE1 and CIRCE2 generators~\cite{Ohl:1996fi} to the
main WHIZARD program. The main core in version 1.xx connects the
different parts of the program via glueing shell scripts that steer
the compilation of different processes as well as the integration,
event generation and the built-in graphical analysis of WHIZARD.  

WHIZARD has been extensively used for linear collider physics, e.g for
the development of the TESLA Technical Design
Report~\cite{Richard:2001qm,AguilarSaavedra:2001rg,Behnke:2001qq}.
The big SLAC event samples for the Standard Model backgrounds have
been generated with WHIZARD. Though here WHIZARD has been used as a
generator for SM backgrounds one of the main focuses of the tool has
always been the realm of beyond the Standard Model (BSM) physics. Many
of these developments have already been present in the legacy branch
WHIZARD 1.xx, but I will summarize them together with the overview of
the new features in WHIZARD 2. 


\section{WHIZARD 2: (New) Technical and Physics Features}

\subsection{Structure and technical features}

\begin{figure}
  \centering
  \includegraphics[width=.8\textwidth]{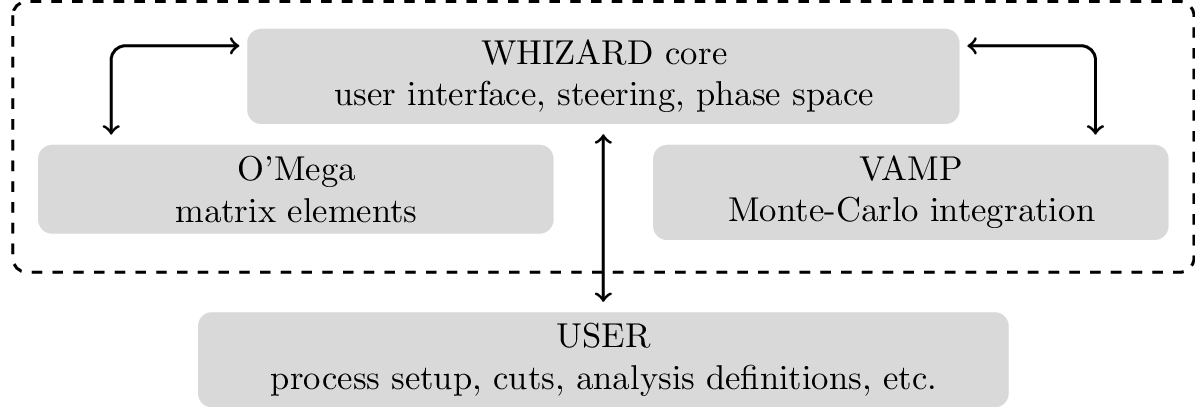}
  \caption{Structure of the WHIZARD program.}
  \label{fig:whizstruct}
\end{figure}

WHIZARD has been basically rewritten since 2007. One of the main
motivations was the inclusion of several features like event-dependent
scales, running couplings, parton showering, handling of a large
number of BSM models which are necessary for the purpose of simulating
signals and backgrounds at the Large Hadron Collider (LHC). But it was
also a question of maintenance of the code, documentation, easening
release productions, bug fixes, and treating regressions that made a
complete rewriting of the code necessary. Since roughly the same time
WHIZARD is located at the HepForge web page, \cite{hepforge}, where
also the revision control system of the project has been moved to. 
With the start of the first release candidates of WHIZARD 2 late in
2009 the line of development for the legacy branch, 1.xx, stopped with
revision 1.94. Until then, only bug fixes and documentation issues are
tackled, and the latest release, 1.97, appeared with a completed
manual documenting the final status and usage of WHIZARD 1.9x. For the
new release branch, the version system has changed to the triple number
system, i.e. (main release).(major version).(minor version). The first
release was in April 2010 for the MC4BSM workshop in Copenhagen, the
actual release at the moment is 2.0.7 from March 2012. 

WHIZARD 2 now is a well-structured program containing the exclusive
optimized matrix element generator O'Mega~\cite{Moretti:2001zz,O2}, the
multi-channel adaptive integration package VAMP~\cite{Ohl:1998jn}, the
two programs CIRCE1 and CIRCE2~\cite{Ohl:1996fi} for ISR,
beamstrahlung and photon collider physics, as well as tools for
graphical data analysis. The basic structure of the program is shown
in Fig.~\ref{fig:whizstruct}. The rewriting of the code (in total more
than 60,000 lines of new code) was a major undertaking. The code has
been completely streamlined, in the sense that now there are only
programming languages used, Fortran2003 and OCaml (for the matrix
element generator O'Mega). All system calls to binaries are done from
the Fortran code itself, so that all shell and Perl scripts have been
abandoned. A huge standardization of modern programming tools was the
usage of the {\em autotools}, i.e. {\em automake/autoconf/libtool}
setup which leads to a much easier control of distributions and easier
maintenance (e.g. regressions etc.). To further control the line of
development, the revision control system (subversion) at the HepForge
page is used, together with the trac system for bug, feature request
and enhancement tickets for the project management of the software. A
cruise control system is used which checks new submissions to the
software repository for compatibility for different compiler suites
and operating systems and runs a very large class of compatibility
tests, sanity and regression checks. A very clean modularization has
been achieved using the object-oriented features in
Fortran2003. WHIZARD 2 now works as a shared library, which makes a
core re-compilation unnecessary whenever one physics process had been
changed. New processes can be dynamically included, while the old
static option is still available, e.g. for the use in batch systems
and on the Grid. The matrix elements which for LHC multi-leg processes
can become rather lengthy are automatically split up in subroutines
which makes compilation by over-eager compiler optimizers much
faster. WHIZARD can also be run as a shell (WHISH) now, though this is
still in an experimental status. For using parallelization and
multiple threads, an OpenMP parallelization for the helicity
amplitudes has been set up, while an MPI parallelization of the
multi-channel integration will be released soon.   

\begin{figure}
  \begin{center}
  
\begin{verbatim}
                   cuts = any 5 degree < Theta < 175 degree
                            [select if abs (Eta) < eta_cut [lepton]]
                   cuts = any E > 2 * mW [extract index 2 
                                            [sort by Pt [lepton]]]
\end{verbatim}
  \end{center}
  \caption{Example for a SINDARIN scripting language expression for cuts.}
  \label{fig:sindarin}
\end{figure}

The program can be downloaded from the HepForge page, unpacked and
then the standard steps should be taken to compile and install it: 
{\em configure}, {\em make}, and {\em make install}. For the
configuration, it might be necessary to specify paths or flags for
external programs to be linked in, like e.g. LHAPDF, StdHEP, HepMC.
Before the last {\em make install} step, an optional {\em make check}
is recommended to ascertain that everything runs correctly on the
current system. WHIZARD 2 is intended to be installed centrally,
e.g. in {\em usr/local} but can also be installed locally without
administrator rights. Each user can then work in his own home or work
directory. 


\subsection{Physics and Performance features}

In this section I summarize the main physics features of WHIZARD with
a special emphasis both on the new developments in WHIZARD 2 as well
as on the ILC-relevant features. First of all, there was an
improvement on the already quite performant phase space setup of
WHIZARD, where due to a symmetrized phase space forest construction a
further performance gain could be achieved. The new modular structure
of WHIZARD 2 made it possible to easily include event-dependent scales
like they are used in parton density functions (PDFs) as well as
running coupling constants like $\alpha_s$. A very powerful invention
was the new steering syntax of the program, Scripting INtegration,
Data Analysis, Results display and INterfaces, or short SINDARIN. This
is similar to a scripting language, and allows to easily define
arbitrary (algebraic) expressions for cuts, scales etc. as well as to
denote all the commands necessary to generate matrix elements, compile
them, integrate them, generate events and set up an
analysis. Fig.~\ref{fig:sindarin} shows an example for a cut
definition in SINDARIN: the first line selects any lepton with polar
angle 5 degrees away from the beam axis under the condition that its
absolute rapidity is below some predefined cut variable. The second
line selects any second-hardest lepton in $p_T$ if its energy exceeds
twice the $W$ mass. Analysis expressions and histograms can be defined
in the same way. 

\begin{figure}
  \centering
  \includegraphics[width=.35\textwidth]{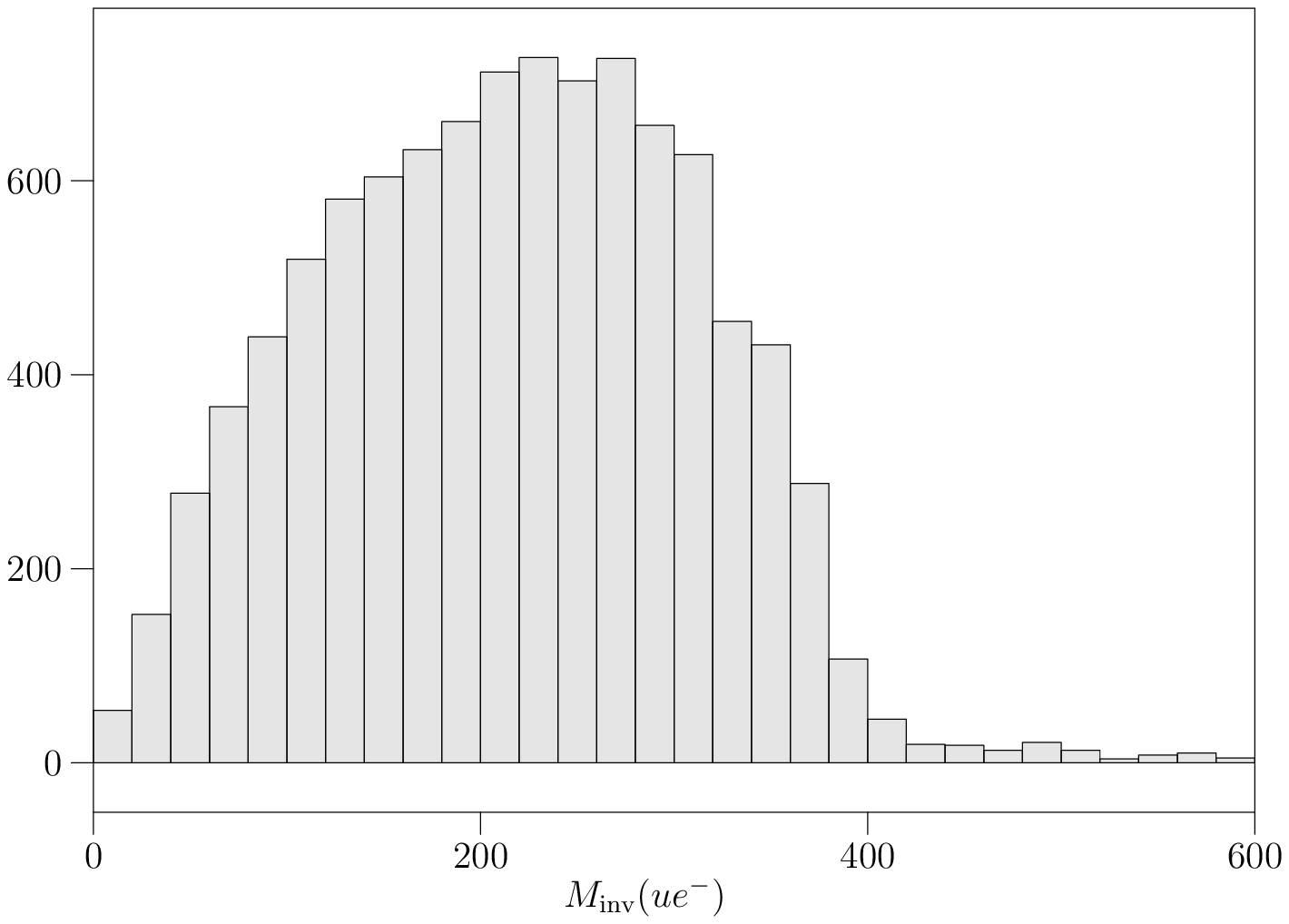}
  \includegraphics[width=.35\textwidth]{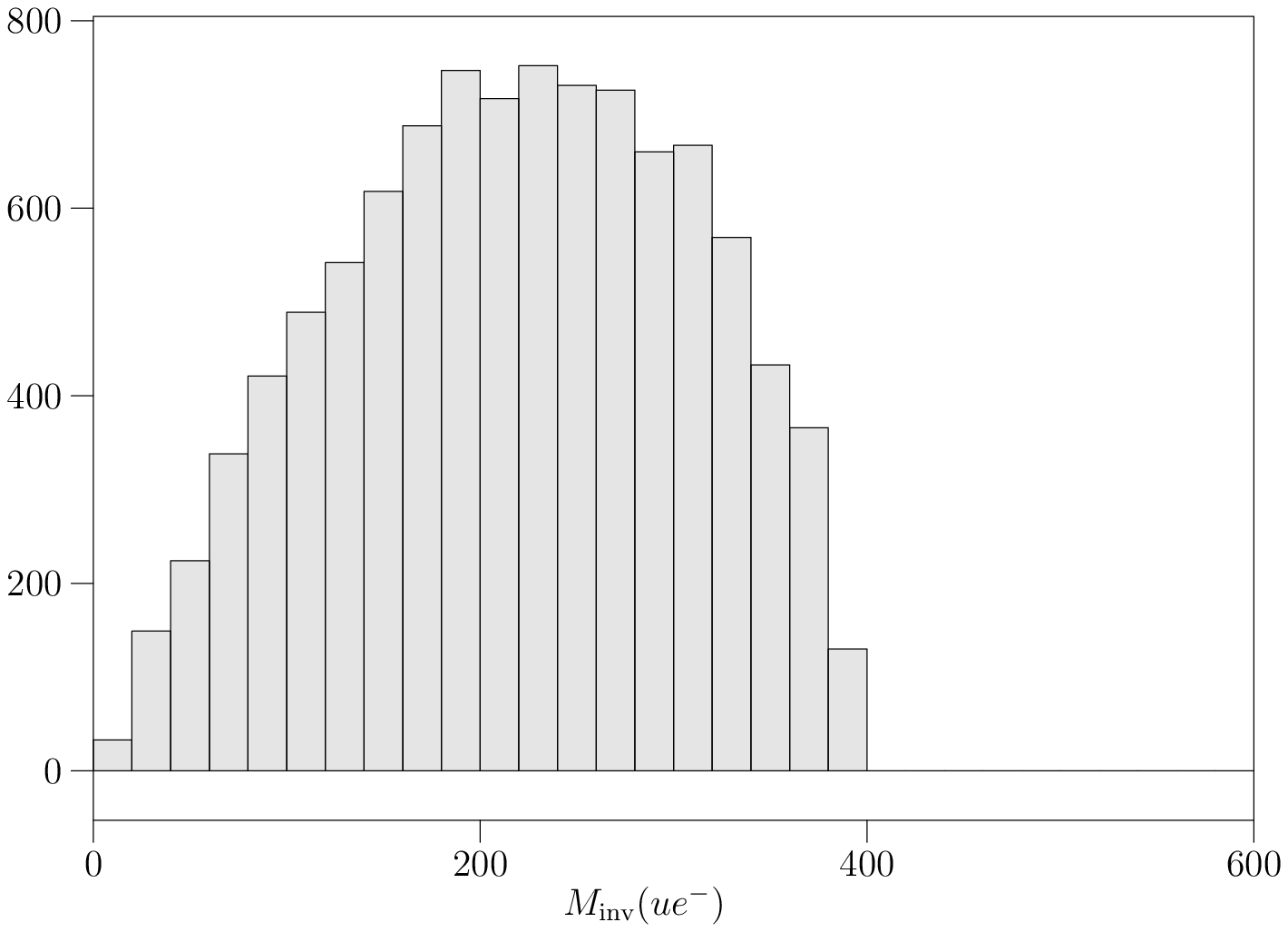} \\
  \includegraphics[width=.35\textwidth]{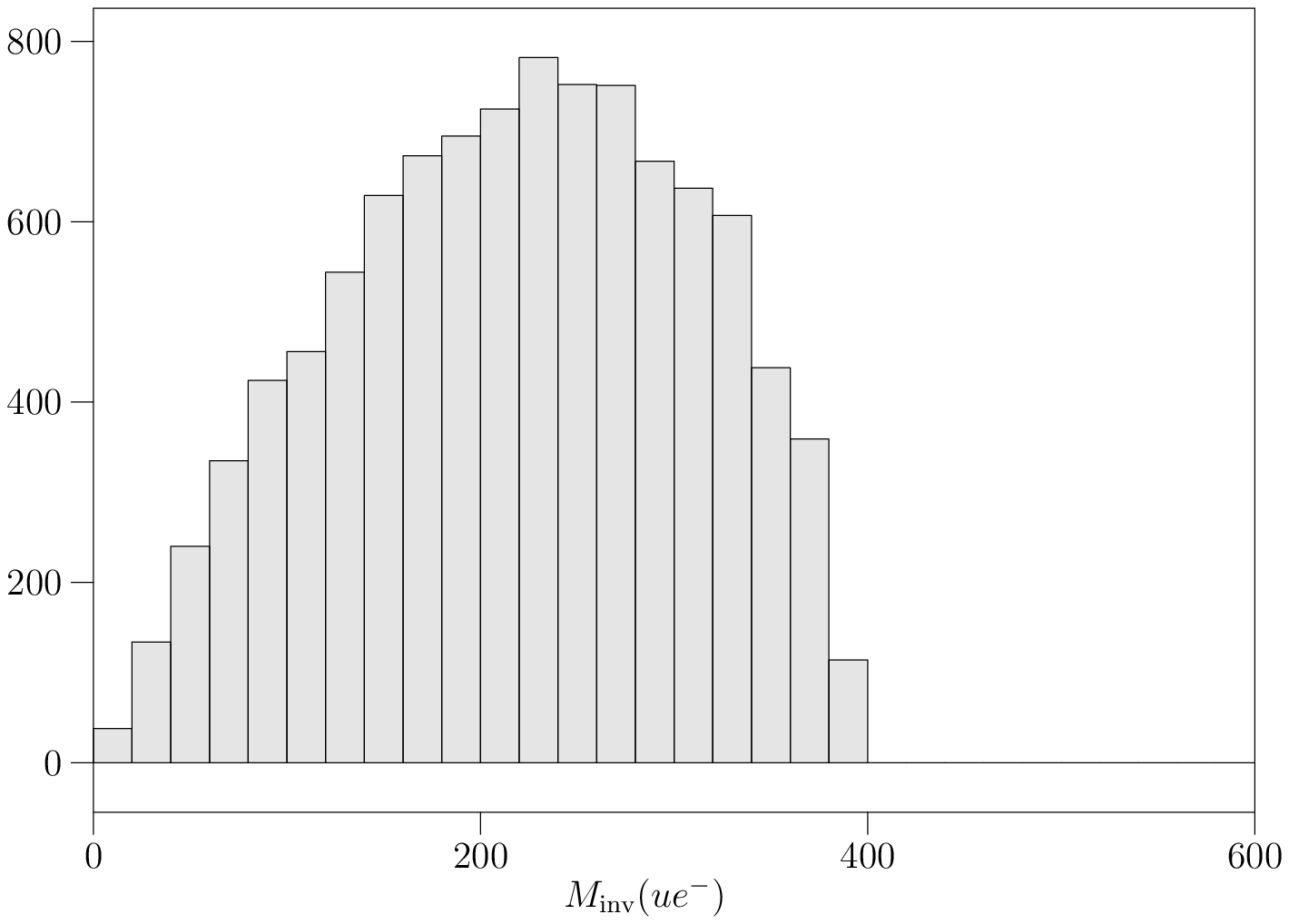}
  \includegraphics[width=.35\textwidth]{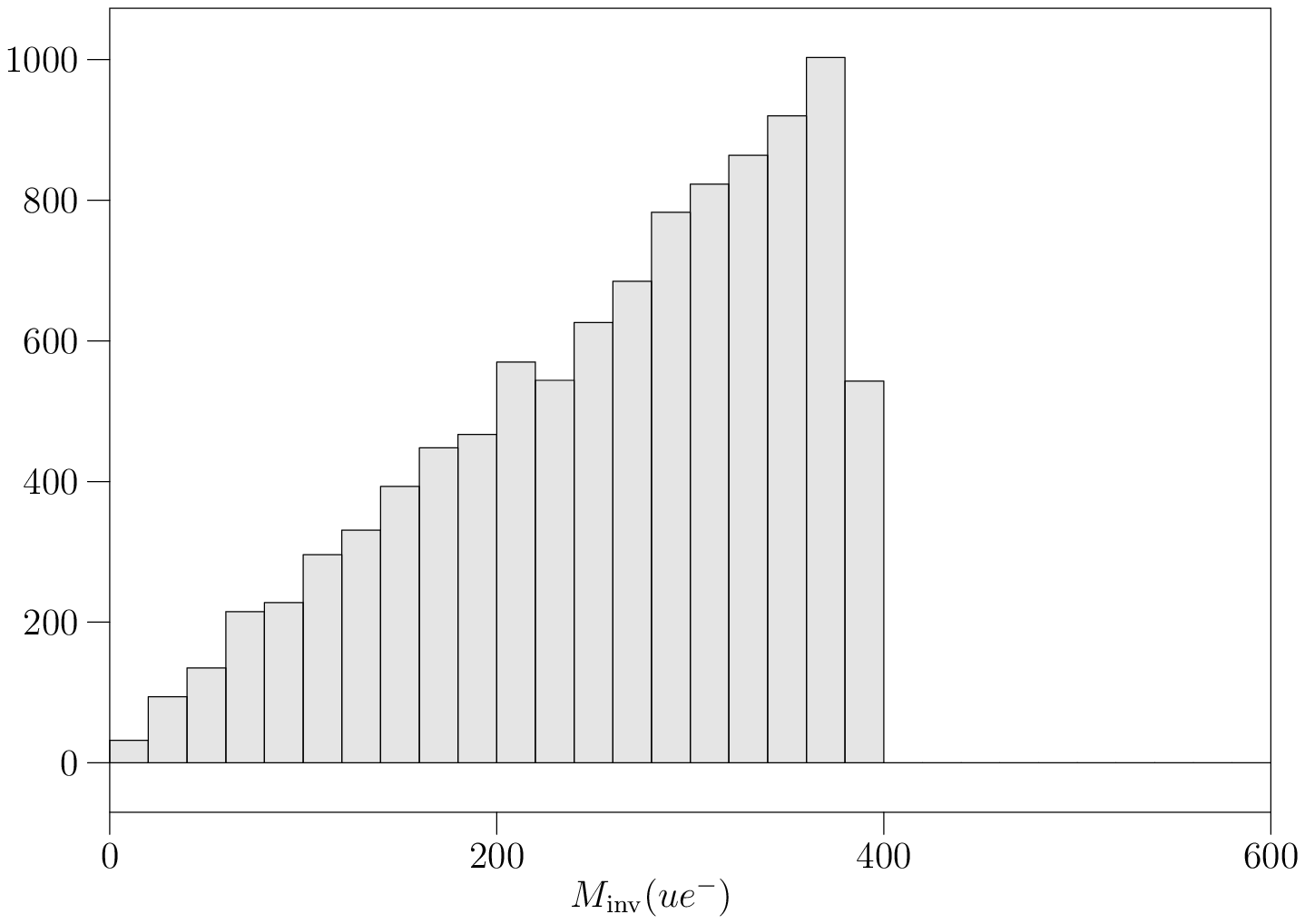} 

  \caption{Factorization of processes in distributions: the jet-lepton
    invariant mass is shown for squark pair production at the LHC,
    where one of the squarks subsequently decays into a jet, a lepton
    and the lightest neutralino. Upper left: full matrix element,
    upper right factorized with full spin correlations, lower row:
    factorized with classical (left) and no spin correlations (right),
    respectively. } 
  \label{fig:cascade}
\end{figure}

WHIZARD 2 uses process libraries, which allows the usage of processes
from different BSM models in parallel. As not the multi-leg matrix
elements, but the high-dimensional phase space integration is the
major bottleneck for going to higher and higher multiplicities,
factorizing amplitudes into production and subsequent decays is (in a
well-defined approximation) not to bad an idea. WHIZARD 2 realizes
this for the event generation and hence distributions where the user
can specify whether he wants no spin correlations, only classical spin
correlations (i.e. the diagonal of the spin density matrix) or full
spin correlations. An example for squark pair production where one of
the two squarks decays via a slepton into jet, lepton and the lightest
neutralino is shown in Fig.~\ref{fig:cascade}. One is able to define
containers of particles for decays and can therefore handle inclusive
processes and decays. With respect to WHIZARD 1.xx, the algorithms for
the flavor sums of initial and final state particles have been greatly
improved. A more elaborate elimination of redundancies from summation
over internal and external combinations of flavours (particularly
quarks in jets, especially for LHC physics) will be available soon and
is expected to further improve both code size and speed. For the
analysis, the graphical package GAMELAN based on LaTeX and MetaPost
has been also improved. Again on the technical side, the algorithm
using MD5 check sums has been revisited, such that is now possible to
reuse every bit and piece of the steps: the code, the object files,
the phase space setup file, the integration grids and the already
generated events, whenever those things are still compatible with the
setup in the input file. Other new features, that will be discussed in
more detail below, are the interface~\cite{Christensen:2010wz} to the
program FeynRules~\cite{Christensen:2008py} which allows to include a
new BSM model just by specifying its Lagrangian, and the initial and
final state parton shower of WHIZARD~\cite{Kilian:2011ka} together
with an MLM matching procedure between hard matrix elements and the
parton shower. 

\subsection{Fields, Beams, Interactions, Models in WHIZARD}

In the discussion of the implemented physics content in WHIZARD (2),
we first start with the hard matrix elements, particle types,
interaction types, Lorentz structures etc. The possible particle types
in WHIZARD contain scalars, spin 1/2 fermions (both Dirac and
Majorana) together with fermion-number violating vertices following
the rules in~\cite{Denner:1992me,Denner:1992vza}, spin 1 particles
(both massless and massive, in unitarity and Feynman gauge as well as
in principle for arbitrary $R_\xi$ gauges), spin 3/2 particles (only
as Majorana particles in their incarnation as gravitinos), as well as
spin 2 particles (massless and massive). Particles could be dynamic
(i.e. propagating particles) or pure insertions. The latter can
e.g. be used as spurion fields in operator insertions. There are also
unphysical particles for testing purposes inside Ward- and
Slavnoy-Taylor identities (see e.g.~\cite{Reuter:2002gn,Ohl:2002jp}). 
Note that for all the particle types there are routines that add up to
a large test suite, testing (especially numerically) equations of
motion, transversality, irreducibility of the on-shell fields as well
as e.g. Majorana proerties of different vertices. 

For the vertices, there is a huge list of Lorentz structures that are
supported by WHIZARD ranging from purely scalar couplings over
scalar-vector couplings (incl. dimension 5 operators), pure vector
couplings, fermionic couplings to scalars, to vectors, to tensors as
well as dimension 5 and 6 operators that appear e.g. in the context of
supersymmetric Ward identities), as well as gravitino couplings of
dimensions 5 and 6. Completely general Lorentz structures that will
allow an automatic generation of a library with the corresponding
Fortran routines is under construction, and will presumably be ready
by the end of the year. 

Color flows in WHIZARD are generated in the color flow
formalism~\cite{Maltoni:2002mq,colorflow}. While in the legacy version
WHIZARD 1.9x this was done in a rather slow approach with the help of
a PERL script, in WHIZARD 2 this was performed directly inside the
core of O'Mega and finally even more refined as a colorizing of the
Directed Acyclical Graph (DAG) as a representation of the colored
amplitude~\cite{O2}. Though in principle every $SU(N)$ gauge group is
supported, we focus here on standard $SU(3)$ for QCD. At the moment,
the fundamental and anti-fundamental representations are supported,
the adjoint representation, which already covers all standard
particles in the SM, SUSY and extra-dimensional models. In preparation
are generalized color structures including color sextets and decuplets
as well as  baryon-number violating vertices as in $\epsilon_{ijk}
\phi_i \phi_j \phi_k$. These color structures are foreseen to be made
public in late 2012. 

\begin{figure}
  \centering
  \begin{verbatim}
              beams = p, p => lhapdf { $lhapdf_file = "cteq5l.LHgrid" }
              beams = p, p => pdf_builtin {$pdf_builtin_set = "mstw2008nlo"}
              beams = e1, E1 => circe1 => isr
              beams = A, A => circe2 { $circe2_file = "teslagg_500.circe" }
              beams = e1, E1 
                   => beam_events { $beam_events_file = "uniform_spread_2.5%.dat" }
              beams = e1, E1 => user_strfun ("escan"), none
\end{verbatim}
  
  \caption{Structured beams in WHIZARD 2 as they appear in SINDARIN commands.}
  \label{fig:structbeams}
\end{figure}

Concerning structured beams, the complete setup for beam structure
relevant for lepton colliders from WHIZARD 1 have been taken over in
or re-implemented for WHIZARD 2, while the support for structured
beams for hadron colliders has been much enlarged and
modernized. Examples for structured beams as SINDARIN commands are
shown in Fig.~\ref{fig:structbeams}~\footnote{For a complete overview
  of SINDARIN commands, cf. the WHIZARD manual~\cite{manual}.}. For
lepton colliders, initial state radiation (ISR) is implemented
according to the calculations presented
in~\cite{Skrzypek:1990qs,Jadach:1990vz} which contains the resummed
results for soft-collinear photon
from~\cite{Gribov:1972rt,Kuraev:1985hb} together with the explicit
calculation of hard-collinear photons up to third order in
perturbation theory. WHIZARD can also generate explicity the $p_T$
distribution of the photons in the event as well as of the electron
beam remnants from the ISR recoil. 

Polarized beams are supported, where it is possible to specify
arbitrary polarization states (not only linear or circular
polarization modes) by using an explicit spin density matrix as
input. Beamstrahlung, i.e. the deformation of the beam spectrum due to
macrosopic (classical) electromagnetic interactions can be simulated
via the CIRCE module~\cite{Ohl:1996fi}, while photon collider spectra
from Compton back scattering are contained in the CIRCE2 generator
within the WHIZARD package. External beam spectra which are basically
long lists of energy ratio (or explicit energy) values can be read in,
or user-defined code can be included, compiled and linked in the
dynamic setup of WHIZARD 2 at runtime. What is at the moment not (yet)
implemented is electromagnetic final state radiation (FSR) using a
Yennie-Frautschi-Suura approach~\cite{Yennie:1961ad}. 

Concerning hadronic beam environments, the support for PDFLIB inside
CERNLIB for PDFs has been abandoned in WHIZARD 2. Like WHIZARD 1,
WHIZARD 2 now exclusively contains an interface to the LHAPDF external
library~\cite{Whalley:2005nh} supporting in principle all (modern) PDF
sets, including photon PDF and pion PDFs. To be independent from
installing LHAPDF and linking it into WHIZARD, the most prominent and
recent PDF sets have been directly included into WHIZARD together with
the routines for the running strong coupling from the PDF
collaborations. Hadronization as well as hadronic events can be
simulated through PYTHIA~\cite{Sjostrand:2001yu} which ships with the
main WHIZARD distribution. Of course, it also possible to write out parton
level events into some event file, read them in into a different
hadronization program and then read the hadronic event file back into
WHIZARD for an analysis. 

\subsection{Parton Shower}

\begin{figure}
  \centering
  
  \includegraphics[scale=.43,angle=-90]{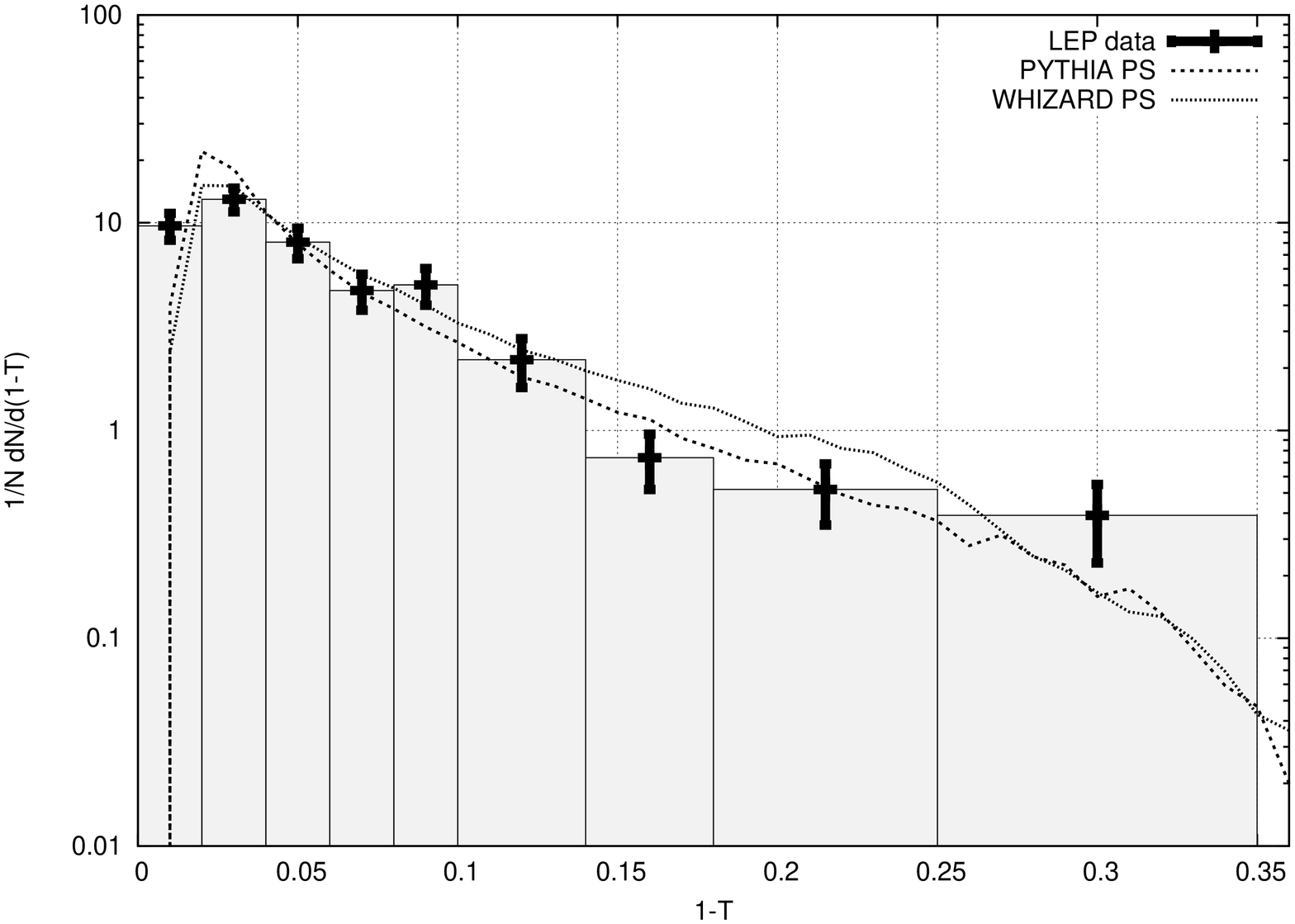}
  \includegraphics[scale=.43, angle=-90]{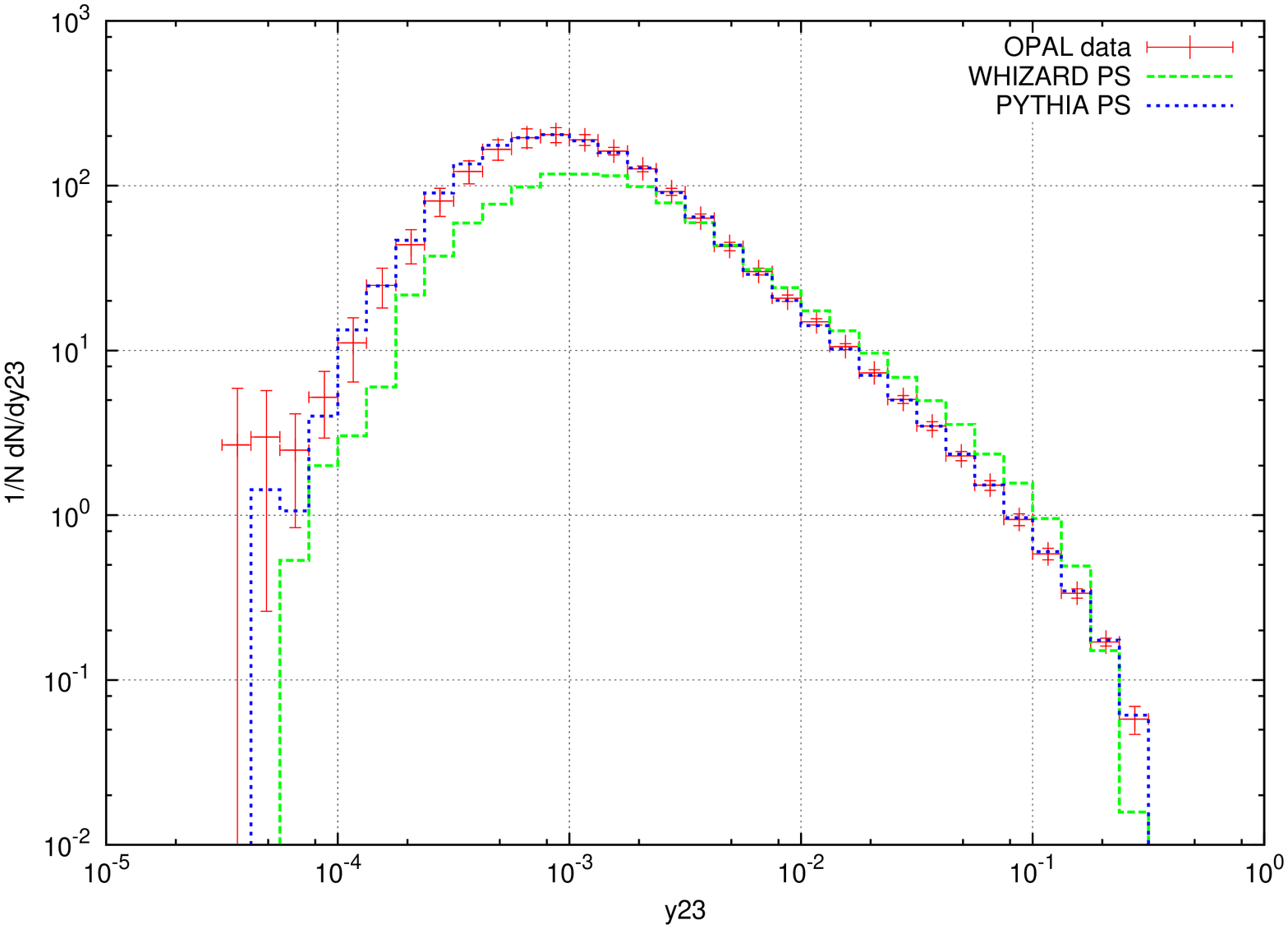}

  \caption{Validation of the analytic final state shower in WHIZARD:
    On top the thrust distribution in $e^+ e^- \to$ jets, where
    the grey histograms with error bars are the data, the dashed line
    is PYTHIA, the dotted WHIZARD, on the bottom the $Y_{23}$ parameter
    which shows the value of the jet definition parameter at which a
    two-jet event starts to be resolved as a three-jet event (in red
    [error bars] OPAL data, the green curve shows WHIZARD, while the
    blue one os PYTHIA). For more details cf.~\cite{Kilian:2011ka}.} 
  \label{fig:shower}
\end{figure}

Parton showering can be done as in WHIZARD 1 with an external program
that is either linked to WHIZARD or via the pipe over an external
event file that is to be converted from partonic to hadron level. In
WHIZARD 2, the latest Fortran version of
PYTHIA~\cite{Sjostrand:2001yu} is included in the distribution
tarball, so parton showering via PYTHIA (like hadronzation and
hadronic decays) can be directly steered from the SINDARIN input
file. In WHIZARD 2, there are now two homebrew parton showers, one
along the lines of the PYTHIA parton shower as $k_T$-ordered shower
including angular ordering, the other one an analytic parton
shower. The details of this latter shower are described in full detail
in~\cite{Kilian:2011ka} (also cf. references therein). Concerning the
original analytic final state parton shower~\cite{Bauer:2008qh},
several improvements have been made, like a running scale of the
strong coupling constant and color coherence by imposing angular
ordering. A comparison to experimental results from the LEP
collaborations have been made, cf. Fig.~\ref{fig:shower}. The main new
feature, however, is the analytic initial state shower, which is the
main part of~\cite{Kilian:2011ka}. There, an automatic MLM-type
matching procedure~\cite{MLM} has been implemented to smoothly connect
high-$p_T$ tails of jet distributions e.g. in Drell-Yan processes with
the low-$p_T$ regime. As this is a workshop on linear collider, I do
not go into the details of the initial state shower, which has been
compared to Tevatron and LHC data in~\cite{Kilian:2011ka}, here.

\begin{table}
  \centering
  
        \begin{center}
          \begin{tabular}{|l|l|l|}
            \hline
            MODEL TYPE & with CKM matrix & trivial CKM \\
            \hline\hline
            QED with $e,\mu,\tau,\gamma$ & -- &  \tt{QED} \\
            QCD with $d,u,s,c,b,t,g$ & -- &  \tt{QCD} \\
            Standard Model        & \tt{SM\_CKM} & \tt{SM} \\
            SM with anomalous gauge couplings &  \tt{SM\_ac\_CKM} &
            \tt{SM\_ac} \\
            SM with anomalous top couplings &  \tt{SMtop\_CKM} &
            \tt{SMtop} \\
            SM with K matrix &  --- &
            \tt{SM\_KM} \\\hline
            MSSM &   \tt{MSSM\_CKM} & \tt{MSSM} \\
            \hline
            MSSM with gravitinos &   --- & \tt{MSSM\_Grav} \\
            \hline
            NMSSM &   \tt{NMSSM\_CKM} &
            \tt{NMSSM} \\
            \hline
            extended SUSY models &   --- & \tt{PSSSM} \\
            \hline
            Littlest Higgs &  --- & \tt{Littlest} \\
            \hline
            Littlest Higgs with ungauged $U(1)$ &  --- &
            \tt{Littlest\_Eta} \\
            \hline
            Littlest Higgs with $T$ parity &  --- &
            \tt{Littlest\_Tpar} \\
            \hline
            Simplest Little Higgs (anomaly-free) &  --- &
            \tt{Simplest} \\
            \hline
            Simplest Little Higgs (universal) &  --- &
            \tt{Simplest\_univ} \\
            \hline
            3-site model &  --- &
            \tt{Threeshl} \\
            \hline
            UED & --- & \tt{UED} \\
            \hline
            SUSY Xdim. (inoff.) & --- & \tt{SED} \\
            \hline
            SM with $Z'$ & --- & \tt{Zprime} \\
            \hline
            SM with gravitino and photino & --- & \tt{GravTest} \\
            \hline
            Augmentable SM template & --- & \tt{Template} \\
            \hline
          \end{tabular}
        \end{center}
  \caption{List of implemented BSM models in WHIZARD.}
  \label{tab:bsm}
\end{table}

\subsection{Models and BSM physics}

Coming back to hard matrix elements, many BSM models have been
implemented in WHIZARD and used for LHC and ILC simulations, and most
of them have been validated with the help of the FeynRules interface
of WHIZARD. Among these are, first of all, SUSY
models~\cite{Hagiwara:2005wg,Kalinowski:2008fk} have been implemented,
the MSSM together with implementations of non-minimal
models like the NMSSM~\cite{Reuter:2009ex} or extended SUSY 
models~\cite{Kilian:2006hh,Braam:2010sy,Reuter:2010nx}. Already in
WHIZARD 1 existed an interface to other codes following the SUSY Les
Houches Accord (SLHA
1/2)~\cite{Skands:2003cj,Allanach:2008qq,AguilarSaavedra:2005pw}. 
Also some pioneering work on the combination of SUSY NLO matrix
elements with the electromagnetic showers have been
done~\cite{Robens:2008sa,Kilian:2006cj}. A second focus lay on Little
Higgs models (with and without T-parity), again with several studies
for linear collider physics~\cite{Kilian:2004pp,Kilian:2006eh}. 
On the more exotic side, models based on noncommutative spacetime have
been studied with
WHIZARD~\cite{Ohl:2004tn,Alboteanu:2006hh,Ohl:2010zf}. 
One of the original motivations was the study on a strongly
interacting sector of electroweak symmetry breaking, which has been
pursued in WHIZARD 2 both along the lines of anomalous
couplings~\cite{Beyer:2006hx} as well as in terms of new resonances in
the electroweak sector~\cite{Alboteanu:2008my}. For the unitarization
of these channels, a method had to be found to distinguish in the
framework of the DAGs of the matrix element generation s- from
t-/u-like channels. 

Table~\ref{tab:bsm} gives a list of all the models that
are implemented. For implementing a new model, it is highly
recommended that this is done via the WHIZARD-FeynRules
interface~\cite{Christensen:2010wz}.

\subsection{NLO development in WHIZARD}

There has been some work on the inclusion of (virtual) NLO corrections
into WHIZARD mentioned in the previous paragraph in the context of
SUSY studies at the ILC. The goal of the more recent developments is
to have a setup for NLO calculations and simulations within WHIZARD
for both LHC and ILC physics that is as automated as possible. NLO
calculations nowadays are mostly based on some sort of subtraction
formalism, that groups the soft-collinear divergences into specific
parts of the calculations to make them finite and performable for a
phase-space integration. The most widely used is the Catani-Seymour
dipole subtraction formalism~\cite{Catani:1996vz,Catani:2002hc}.
A first proof-of-principle implementation of the integrated and
unintegrated dipoles have been done
in~\cite{Binoth:2009rv,Greiner:2011mp}. An automated generation of the
CS dipoles is in construction at the moment, but already gives correct
results for QED processes. Along with the dipoles comes an
implementation of using several instances of the process setup within
the phase space integration, which is necessary for the unintegrated
dipoles in order to take care of the squeezed kinematics in the phase
space integration of the subtraction terms. The implementation will be
made public several steps (together with a BLHA
interface~\cite{Binoth:2010xt}) several steps from summer until the
end of this year.  

\section{Summary and Outlook}

WHIZARD 2 is a completely newly structured update of an already
versatile multi-purpose Monte Carlo event generator that has been
released with many new technical and physics features in April 2010. 
Many further improvements and features have been added in the past two
years. Though the main motivation for the restructuring of the code
was to deal with the complexities of LHC physics, linear collider
physics has always been a major field of application for
WHIZARD. Quite recently, all relevant features regarding ILC/CLIC
physics from WHIZARD 1 have been reimplemented in WHIZARD 2, and many
improvements on the phase space setup, color, parton shower, BSM
models, speed and performance, maintenance and usability have been
made. Continuos effort will go specifically into the direction of
multi-leg amplitudes, NLO development and more BSM coverage to be
ready for the high-energy phase of LHC and a possible future linear
collider. 

\section{Acknowledgments}

JRR thanks the WHIZARD team, F. Bach, H.-W. Boschmann, W. Kilian,
T. Ohl, S. Schmidt, C. Speckner, M. Trudewind, D. Wiesler, and
T. Wirtz as well as F. Braam for the joint effort to make this project
successful.  

This project has been partially supported by the Helmholtz Alliance
``Physics at the Terascale'', the German ministry BMBF, the ministry 
MWK of the German state Baden-W\"urttemberg, the German Research
Association DFG, as well as the Scottish Universities Physics Alliance
(SUPA). JRR wants to thank the institutes at Carleton University,
Ottawa, the Institute of Physics in Freiburg as well as the School of
Physics and Astronomy in Edinburgh as well as the Aspen Center of
Physics for their hospitality, as part of this project has been
realized there. 

 

\begin{footnotesize}

\end{footnotesize}

\end{document}